\documentclass[a4paper,10pt]{article}

\usepackage[utf8]{inputenc}
\usepackage{amsmath, amsthm, amssymb}
\usepackage{geometry}
\usepackage{graphicx}
\usepackage{hyperref}
\usepackage{tikz-cd}
\usepackage{relsize}
\usepackage{algorithmicx}
\usepackage{algorithm}
\usepackage{algpseudocode}
\usepackage{natbib}
\usepackage{float} 



\newtheorem{theorem}{Theorem}

\newtheorem{proposition}[theorem]{Proposition}

\theoremstyle{definition}
\newtheorem{definition}[theorem]{Definition}
\newtheorem{example}[theorem]{Example}

\newcommand{\PP}{\mathbb{P}}
\newcommand{\EE}{\mathbb{E}}
\newcommand{\cov}{\mathrm{Cov}}
\newcommand{\Aa}{\mathcal{A}}
\newcommand{\Cc}{\mathcal{C}}

\newcommand{\NN}{\mathbb{N}}
\newcommand{\PSet}{\mathrm{PSet}}
\newcommand{\MSet}{\mathrm{MSet}}

\title{Oracle-free Boltzmann Sampling for Powersets}
\author{Jean C. Peyen\\University of Dundee, DIICSU\\\texttt{jpeyen001@dundee.ac.uk}}

\date{\today}

\begin{document}
\maketitle

\begin{abstract}
We propose an approach for sampling powersets under the Boltzmann distribution in an \emph{oracle-free} way, i.e. without numerically evaluating the associated generating function. Our approach relies on a Poissonised infinite occupancy model and thinning. It yields an explicit sampler for bounded counting sequences and extends under mild growth conditions. We implement the sampler and find runtimes comparable to existing Boltzmann samplers.
\end{abstract}


\section{Introduction}
Boltzmann samplers \cite{Duchon} generate random combinatorial objects $\gamma\in\Cc$ with probability
\begin{equation}
    \PP(\gamma) = \frac{z^{N(\gamma)}}{C(z)},
\end{equation}
where $C$ is the generating function of $\Cc$ and $z$ is tuned to control the output size through
\begin{equation}\label{eq:tuning}
    \EE(N)= \frac{z C'(z)}{C(z)}.
\end{equation}
In the analytic-combinatorics framework \cite{anacomb}, implementing a Boltzmann sampler typically requires an oracle to evaluate $C(z)$ (or related fixed points), often via numerical iteration \cite{Pivoteau}. We call a sampler \emph{oracle-free} if it runs without numerically evaluating $C(z)$ (or any equivalent fixed-point system) at the sampling parameter $z$.

We consider the powerset construction $\Cc=\PSet(\Aa)$ (finite subsets of a class $\Aa$). Existing approaches, such as the parity-filtering method of \cite{Flajolet}, proceed through multisets and inherit an oracle requirement.

In  order to give a direct oracle-free construction, we represent the Boltzmann law on $\PSet(\Aa)$ as the image of a Poissonised infinite occupancy model and recover it by thinning. For bounded counting sequence $(a_n)$ this yields an explicit sampler (Algorithm~\ref{algo:Boltzmann1}) requiring only a bound on $(a_n)$ and uniform sampling from $\Aa_n$. The same template extends to unbounded $(a_n)$ given an explicit dominating intensity (Section~\ref{sec:extension}).

\section{Definitions and notations}
The occupancy problem consists in distributing a given number $m$ of balls among a set of boxes with positive frequencies. The reader may refer to \cite{Feller} for a presentation of the occupancy problem in its classical form or to \cite{Gnedin} for the case with infinitely many boxes.

\begin{definition}
We denote by $\Aa$ a combinatorial class endowed with a size function $N:\Aa \rightarrow \NN_0$ and by $\Cc$ the structure $\PSet (\Aa)$, where the size function is defined by additivity. We define the level sets of the size function and the counting sequence
    \begin{equation}
        \Aa_n = \{\ell \in \Aa~ \vert ~ N(\ell)=n\},~~ a_n = \sharp \Aa_n.
    \end{equation}
\end{definition}
\begin{definition}
We define $\PP$ as the distribution of the occupancy model with a random number of elements $M$. The elements are distributed over $\Aa$ with frequency $f=(f_\ell)_{\ell \in \Aa}$. We generically denote by $\nu=(\nu_\ell)_{\ell\in \Aa}$ the sequence of multiplicities of a configuration with distribution $\PP$. This means that we have
\begin{equation}
    \nu_\ell = \sum_{k=1}^M \mathbf{1}_\ell (X_k),
\end{equation}
where the $X_k$ are i.i.d. with law
\begin{equation}
    \PP(X_k =\ell)=f_\ell.
\end{equation}
\end{definition}

\begin{definition}
We define $\nu^{\prime}$ as the random element of $\Cc$ obtained by setting to $1$ the multiplicity of each element that has a non-zero multiplicity in $\nu$. In particular
\begin{equation}
    \PP(\nu'_\ell =1) = 1-\PP(\nu_\ell=0).
\end{equation}
\end{definition}

\section{Construction of the sampler}\label{sec:theory}
To generate an element of $\PSet(\Aa)$ we first generate a configuration from an occupancy model with boxes labelled by the elements of $\Aa$ and a random number of balls $M$. The distinction between the balls is omitted to obtain an element of $\MSet(\Aa)$. Finally, multiplicities are also omitted to obtain an element of $\PSet(\Aa)$. With an adequate choice of distribution for $M$ and for the occupancy model, this process generates a configuration with the Boltzmann distribution as shown in Proposition \ref{prop:samplingprop}. The algorithm given in Proposition \ref{prop:algo} can be used to implement this construction, as shown in Section \ref{sec:practical}.
\begin{proposition}\label{prop:ProbCov1}
\begin{align}
    \PP(\nu^{\prime}_\ell = 1) &= 1-G_M(1-f_\ell) \label{eq:Prob1}\\
    \cov(\nu^{\prime}_\ell, \nu^{\prime}_{\ell^\prime}) &= G_M(1-(f_\ell+f_{\ell^\prime}))-G_M(1-f_\ell)G_M(1-f_{\ell^\prime}) \label{eq:Cov1}
\end{align}
where $G_M$ is the probability generating function of $M$.
\end{proposition}

\begin{proof}
For \eqref{eq:Prob1},
\begin{equation*}
\PP(\nu^{\prime}_\ell = 1)=1-\PP(\nu_\ell=0)=1-\sum_{m\ge 0}\PP(M=m)(1-f_\ell)^m=1-G_M(1-f_\ell).
\end{equation*}
For \eqref{eq:Cov1}, using inclusion--exclusion,
\begin{equation*}
\PP(\nu^{\prime}_\ell \nu^{\prime}_{\ell^\prime}=1)=1-G_M(1-f_\ell)-G_M(1-f_{\ell'})+G_M(1-(f_\ell+f_{\ell'})),
\end{equation*}
so subtracting $\PP(\nu^{\prime}_\ell = 1)\PP(\nu^{\prime}_{\ell^\prime} = 1)$ gives \eqref{eq:Cov1}.
\end{proof}

\begin{proposition}\label{prop:poisson}~
\begin{enumerate}
    \item  If $M$ follows a Poisson distribution, then the $\nu^{\prime}_\ell$ are mutually independent.
    \item If the $\nu^{\prime}_{\ell}$ are uncorrelated and the $f_\ell$ accumulate at $0$ (i.e. they take infinitely many non-zero values), then $M$ follows a Poisson distribution.
\end{enumerate}
\end{proposition}
\begin{proof}~
\begin{enumerate}
    \item  Assume that $M$ follows the Poisson distribution with parameter $\lambda$. Its probability generating function is $G_M(t) = \exp{} (\lambda (t-1))$ and $\PP(\nu^{\prime}_\ell = 0)=\exp{} (-\lambda f_\ell)$. Consider a subset $E$ of $\Aa$. The probability that $\nu^{\prime}_\ell=0$ for all $\ell\in E$ is
\begin{align*}
    \sum_{m=0}^\infty \PP(M=m)\left(1-\sum_{\ell\in E}f_\ell \right)^m = G_M\left(1-\sum_{\ell\in E}f_\ell \right)= \exp{} \left(-\lambda \sum_{\ell\in E}f_\ell \right) = \prod_{\ell\in E} \exp(-\lambda f_\ell).
\end{align*}
Since this holds for every finite $E\subset\Aa$, the event $\{\nu'_{\ell}=0\ \forall \ell\in E\}=\bigcap_{\ell\in E}\{\nu'_\ell=0\}$ has probability
\[
\PP\bigl(\nu'_{\ell}=0\ \forall \ell\in E\bigr)=\prod_{\ell\in E}\PP(\nu'_\ell=0).
\]
Thus the finite-dimensional distributions factorise, which characterises the joint law of the Bernoulli family $(\nu'_\ell)_{\ell\in\Aa}$. This implies mutual independence of the $\nu'_{\ell}$ by a monotone class argument.

    \item Let $G=G_M$, and let $(p_n)$ be a strictly decreasing sequence of frequencies $p_n=f_{\ell_n}$ such that $p_n\to0$.

For all $n$, define
\begin{equation*}
    H_n(t)=G(1-p_n-t)G(1-p_{n+1})-G(1-p_{n+1}-t)G(1-p_n).
\end{equation*}
Since $1-p_n,1-p_{n+1}\in(0,1)$, the function $H_n$ is analytic in a neighbourhood of $0$. Moreover, for all $k\notin\{n,n+1\}$, Proposition \ref{prop:ProbCov1} gives
\begin{equation*}
    H_n(p_k)=
    G(1-p_n)G(1-p_k)G(1-p_{n+1})
    -G(1-p_{n+1})G(1-p_k)G(1-p_n)
    =0.
\end{equation*}
Since $p_k\to0$, the isolated zeros theorem (see \cite{Rudin}) implies that $H_n$ vanishes identically on the connected component of its domain of analyticity containing $0$.

Let
\begin{equation*}
    d_n=p_n-p_{n+1}>0.
\end{equation*}
For $u\in(0,1-d_n)$, take $t=1-p_n-u$. Then $H_n(t)=0$, hence
\begin{equation*}
    G(u)G(1-p_{n+1})=G(u+d_n)G(1-p_n).
\end{equation*}
Since $G_M$ is strictly positive on $(0,1)$, logarithmic differentiation with respect to $u$ gives
\begin{equation*}
    \frac{G'(u)}{G(u)}=\frac{G'(u+d_n)}{G(u+d_n)}.
\end{equation*}
Let
\begin{equation*}
    r(u)=\frac{G'(u)}{G(u)}.
\end{equation*}
Thus $r(u)=r(u+d_n)$. Since $d_n\to0$, for every fixed
$u\in(0,1)$,
\begin{equation*}
    r'(u)=\lim_{n\to\infty}\frac{r(u+d_n)-r(u)}{d_n}=0.
\end{equation*}

Therefore, $r$ is constant on $(0,1)$. Writing
$r(u)=\lambda$, we obtain
\begin{equation*}
    G'(u)=\lambda G(u).
\end{equation*}
Since \(G(1)=1\), it follows that
\begin{equation*}
    G(u)=\exp(\lambda(u-1)),
\end{equation*}
which is the probability generating function of a Poisson distribution.
\end{enumerate}
\end{proof}

\begin{proposition}\label{prop:samplingprop}
Let $\Aa$ be a non-empty combinatorial structure with size function $N:\Aa\rightarrow \NN_0$. Let $f$ be the distribution
\begin{equation} \label{eq:mult}
    f_\ell = \frac{\ln (1+z^{N(\ell)})}{\ln C(z)}~~,~~ \ell\in \Aa,
\end{equation}
where $C$ is the generating function of $\mathit{\PSet}(\Aa)$,
\begin{equation}
    C(z) = \prod_{\ell\in \Aa} (1+z^{N(\ell)}).
\end{equation}
Suppose that $M$ follows the Poisson distribution with parameter $\lambda = \ln (C(z))$. The strict partition defined by the multiplicities $\nu_\ell^{\prime}$ follows the Boltzmann distribution over $\mathit{\PSet}(\Aa)$.
\end{proposition}

\begin{proof}
    It is an immediate consequence of Proposition \ref{prop:ProbCov1} which gives the marginal law of the $\nu^\prime_\ell$
    \begin{equation*}
        \PP(\nu^\prime_\ell=1)= \frac{z^{N(\ell)}}{1+z^{N(\ell)}}
    \end{equation*}
    and Proposition \ref{prop:poisson} which ensures the independence and allows to recover the Boltzmann distribution. 
\end{proof}
We can avoid the explicit calculation of the parameter $\lambda$ by using the Lewis thinning method. See \cite{Ogata} for a presentation of this method for Poisson processes, from which our case follows.

\begin{proposition}\label{prop:algo}
    
Assume that the counting sequence of $\Aa$ is bounded as follows
    \begin{equation}
        a_n\leq\overline{a},~~\forall n\in \NN_0. 
        \end{equation}
    Then, Algorithm \ref{algo:Boltzmann1} is a Boltzmann sampler for $\Cc$.
    \begin{algorithm}
    \caption{Boltzmann$(z)$}\label{algo:Boltzmann1}
    \begin{algorithmic}[1]
        \State Initialise $\nu' \leftarrow \emptyset$
        \State $\overline{\lambda} \leftarrow \mathlarger{\frac{\overline{a}}{1-z}}$
        \State{$\overline{M} \leftarrow \text{Poiss}(\overline{\lambda})$}
        \For{$i$ from $1$ to $\overline{M}$}
            \State{$n \leftarrow \text{Geom}(1-z)$}
            \If{$\mathlarger{\text{Bern}\left(\frac{a_n}{\overline{a}} \frac{\ln(1+z^n)}{z^n}\right)}$}
                \State{$\ell \leftarrow \text{Unif}(\Aa_n)$}
                \State{$\nu' \leftarrow \nu' \,\cup \, \{\ell\}$}
            \EndIf
        \EndFor
    \end{algorithmic}
    \end{algorithm}
\end{proposition}

\begin{proof}
Interpret $M$ (as in Proposition~\ref{prop:samplingprop}) as the number of arrivals in unit time of independent Poisson processes with rates $a_n\ln(1+z^n)$. Dominating them by processes of rates $\overline{a}z^n$ and thinning yields the acceptance probability $\frac{a_n}{\overline{a}}\frac{\ln(1+z^n)}{z^n}$. The mixture weights are proportional to $\overline{a}z^n$, thus $\PP(n)=z^n(1-z)$ and $n\sim\mathrm{Geom}(1-z)$. Conditional on $n$, objects are sampled uniformly in $\Aa_n$. This is consistent with the conditional uniformity of the Boltzmann distribution.
\end{proof}

Since we assumed that $a_n$ is bounded, sampling from $\text{Unif}(\Aa_n)$ can be done in a bounded time regardless of $n$. In particular, when $\EE(N)$ goes to infinity, the expectation and the variance of the sampling time, both behave as $1/(1-z)$, where $z$ goes to $1$. In general, we cannot give a more accurate statement as $z$ needs to be tuned, either experimentally or through Equation \ref{eq:tuning}.
    
\section{Practical implementation and tests}\label{sec:practical}
We confirm the practical validity of the algorithm described in Proposition \ref{prop:algo} by testing a Python implementation. Although Python is not a high performance language, it facilitates the implementation of this particular algorithm with the $\mathtt{set}$ class, allowing a syntax that remains close to the pseudo-code.

Boltzmann samplers are often used in conjunction with a rejection scheme to control the output size.  We distinguish:
\begin{itemize}
    \item Free samplers, that simply reproduce the Boltzmann distribution without rejection.
    \item Approximate rejection schemes, that repeat the sampler until the size is in a range $[(1-\varepsilon)\,n,(1+\varepsilon)\,n]$.
    \item Exact rejection schemes, that repeat the sampler until the size is exactly equal to a given value $n$.
\end{itemize}
In this section, we report the results of the tests for free and exact sampling schemes. Tests were carried out on an Apple MacBook Air M4. The Python code used to generate the figures is available in the GitHub repository of the project \cite{Repo}. Runtime comparisons are indicative, since the hardware configurations and programming languages differ from the literature references.

\begin{example}\label{ex:plain}
    Consider the classical case of strict partitions, where $\Aa=\NN$. These can be represented graphically with the upper bound of their Young diagram
    \begin{equation*}
        Y(x)= \sum_{n \geq x}\nu'_n, ~~ x\geq 0.
    \end{equation*}
    Here we have $\Aa_n=\{n\}$ and $\overline{a}=a_n=1$. We can give an explicit calibration equation that links the expected size and the parameter $z$
    \begin{equation*}
        z \sim \exp\left(-\frac{1}{\sqrt{c \EE(N)}} \right) ,~~c=\frac{12}{\pi^2}.
    \end{equation*}
    The limit shape after rescaling by the square root of the size is given in \cite{Vershik}
    \begin{equation*}
       e^{\pi y/\sqrt{12}} = 1+e^{-\pi x/\sqrt{12}}.
    \end{equation*}
    Figure \ref{fig:strictpart} confirms that the sampler reproduces this shape as the size goes to infinity (or $z$ goes to $1$).
    
    The expected number of iterations of the \texttt{for} loop is equal to $\overline{M}$, on average it is $\overline{\lambda}$. With the calibration equation we obtain
    \begin{equation*}
        \overline{\lambda} \sim \sqrt{c\EE(N)}.
    \end{equation*}
    Moreover, as shown in Tables \ref{tab:benchstrictapprox} and \ref{tab:benchstrictexact}, and Figure \ref{fig:bench}, runtimes are comparable to those reported in \cite{Flajolet}.
    \begin{figure}[h!]
        \centering
        \begin{minipage}{0.48\textwidth}
            \centering
            \includegraphics[width=\linewidth]{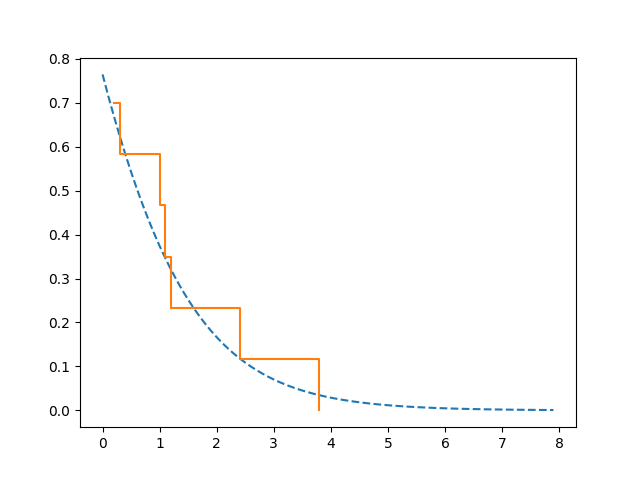}
        \end{minipage}\hfill
        \begin{minipage}{0.48\textwidth}
            \centering
            \includegraphics[width=\linewidth]{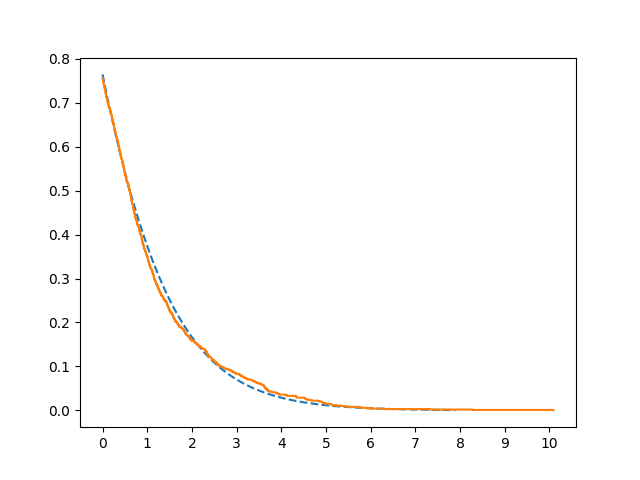}
        \end{minipage}
        \caption{The orange solid line represents a sampled partition, and the blue dotted line represents the scaling limit. On the left panel the partition is of size $100$ and on the right panel it is of size $1\,000\,000$.}
        \label{fig:strictpart}
    \end{figure}
    \begin{table}[h!]
        \centering
        \begin{minipage}{0.49\textwidth}
            \centering
            \begin{tabular}{l|l|l}
                Exp. size  & Avg. time & SD\\
                \hline
                 $10^3$& $3.11 \times 10^{-2}\,ms$  & $2.03 \times 10^{-1}\,ms$ \\
                 $10^6$& $8.33 \times 10^{-1}\,ms$  & $2.00 \times 10^{-1}\,ms$ \\
                 $10^9$& $2.69 \times 10^{1}\,ms$  & $5.34 \times 10^{-1}\,ms$\\
            \end{tabular}
            \caption{Free size sampling (strict partitions)}
            \label{tab:benchstrictapprox}
        \end{minipage}\hfill
        \begin{minipage}{0.49\textwidth}
            \centering
            \begin{tabular}{l|l|l}
                Exp. size  & Avg. time & SD\\
                \hline
                 $10^2$& $1.09\,ms$  & $1.10\,ms$ \\
                 $10^3$& $1.87 \times 10^{1}\,ms$  & $1.88 \times 10^{1}\,ms$ \\
                 $10^4$& $3.77 \times 10^{2}\,ms$  & $3.88 \times 10^{2}\,ms$\\
            \end{tabular}
            \caption{Exact size sampling (strict partitions)}
            \label{tab:benchstrictexact}
        \end{minipage}
    \end{table}
    \begin{figure}[H]
        \centering
        \begin{minipage}{0.48\textwidth}
            \centering
            \includegraphics[width=\linewidth]{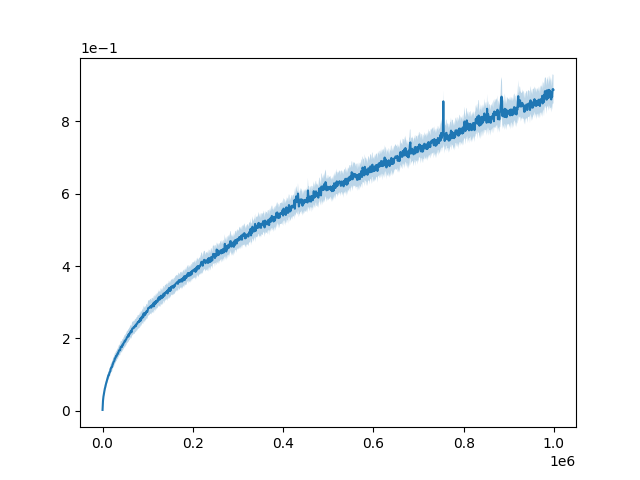}
        \end{minipage}\hfill
        \begin{minipage}{0.48\textwidth}
            \centering
            \includegraphics[width=\linewidth]{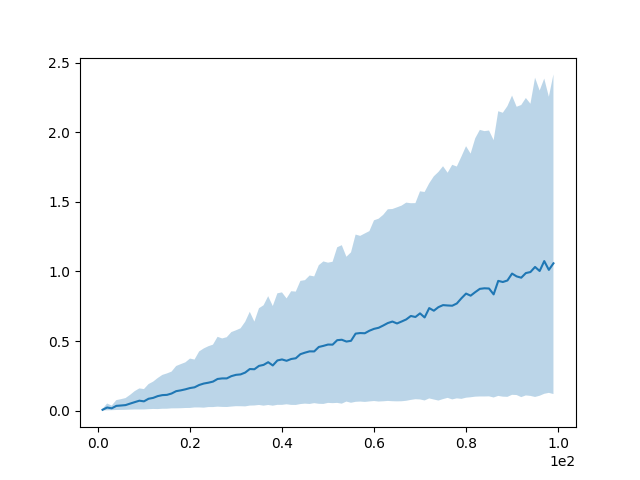}
        \end{minipage}
        \caption{Sampling times (in $ms$) for the free sampler (left panel) and the exact sampler (right panel). The shaded region represents the interval between the percentiles $10$ and $90$.}
        \label{fig:bench}
    \end{figure}
\end{example}

\begin{example}\label{ex:square}
    We consider the case of strict partitions into squares with two parameters, the size and the number of parts. Although it is not covered in Section~\ref{sec:theory}, it is an immediate generalisation and illustrates that the same thinning construction accommodates multi-parameter (e.g. bivariate) Boltzmann sampling with no additional conceptual difficulties. This class of partitions has been treated in \cite{pey2}. The frequency of parts in $\nu$ is given by
    \begin{equation}\label{eq:mult-bivar}
        f_n = \frac{\ln (1+z_2z_1^{n})\mathbf{1}_S(n)}{\ln C(z_1,z_2)}.
    \end{equation}
    where $S$ designates the set of the perfect squares. The counting sequence is the indicator of $S$,
    \begin{equation*}
        a_n = \mathbf{1}_S(n) \leq 1=\overline{a}.
    \end{equation*}
    we have
    \begin{equation*}
        \overline{\lambda}= \frac{z_2}{1-z_1}
    \end{equation*}
    and the acceptance probability to use in the sampler is
    \begin{equation*}
        \mathbf{1}_S(n) \frac{\ln(1+z_2z_1^n)}{z_2z_1^n}.
    \end{equation*}

    In an adequate limit regime, the expected size and length are linked to the parameters $z_1$ and $z_2$ by the formula:
    \begin{equation*}
        z_1 \sim \exp\left(-\frac{\EE(M)}{2\EE(N)} \right),~~ z_2 \sim \sqrt{\frac{\kappa}{2}}\frac{1}{\Gamma(3/2)}, ~~ \kappa = \frac{\EE(M)^3}{\EE(N)}.
    \end{equation*}
    The limit shape is the survival function of the gamma distribution with shape parameter $1/2$ and scale $1$
    \begin{equation*}
        1-\frac{1}{\sqrt{\pi}}\int_0^x u^{-1/2}e^{-u}\, du
    \end{equation*}
    under the rescaling
    \begin{equation*}
        \widetilde{Y}(x) = \frac{1}{\EE(M)}\, Y\left(\frac{2\EE(N)x}{\EE(M)}\right).
    \end{equation*}
    
    \begin{figure}[h!]
    \centering
    \begin{minipage}{0.49\textwidth}
        \centering
        \includegraphics[width=\linewidth]{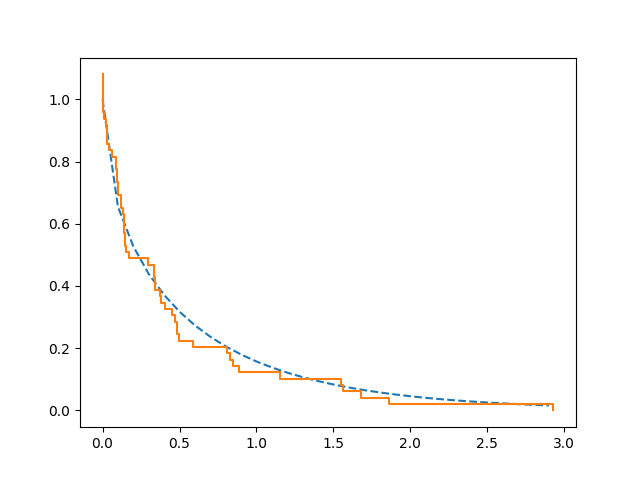}
    \end{minipage}\hfill
    \begin{minipage}{0.49\textwidth}
        \centering
        \includegraphics[width=\linewidth]{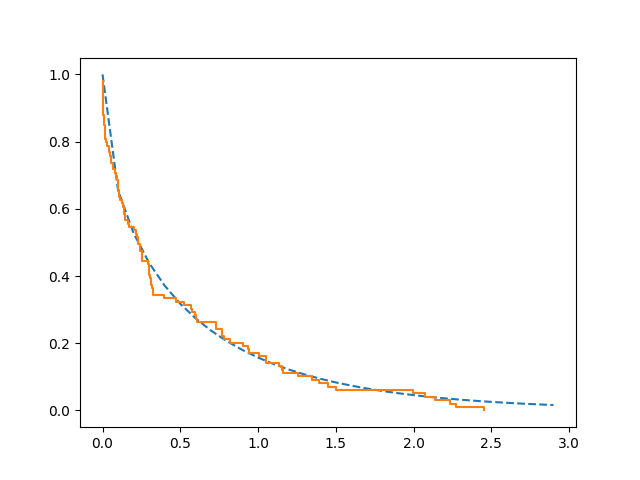}
    \end{minipage}
    \caption{The orange solid line represents a sampled partition, and the blue dotted line represents the scaling limit. On the left panel the partition has been sampled with $\EE(M)=50$, $\EE(N)=10^9$ and on the right panel it has been sampled with $\EE(M)=100$, $\EE(N)=10^{12}$.}
    \label{fig:strictpart-squares}
    \end{figure}
    We can show that the expected number of iterations of the loop satisfies the following
    \begin{equation*}
        \overline{\lambda} \sim \frac{\sqrt{2 \EE (N) \EE(M)}}{\Gamma(3/2)}.
    \end{equation*}
    Actual execution times are reported in Table \ref{tab:benchstrictsquare}. The reported performance are sensibly similar to those of \cite{pey2}. As a sanity check, we can run the code to show that the sampling time is of the order of $0.1 \, ms$ for $\EE(N)=12\,500$, $\EE(M)=5$.

    \begin{table}[h!]
    \centering
    \begin{tabular}{l|l|l|l}
        Expected size  & Expected length & Average Sampling Time & Sampling Time Standard Deviation\\
        \hline
            $10^6$& $5$ & $1.18\,ms$  & $3.16 \times 10^{-2}\,ms$ \\
            $10^6$& $10$ & $1.74\,ms$  & $4.10 \times 10^{-2}\,ms$ \\
            $10^6$& $15$ & $2.07\,ms$  & $5.28 \times 10^{-2}\,ms$ \\
            $10^6$& $20$ & $2.41\,ms$  & $5.39 \times 10^{-2}\,ms$ \\
        \hline
            $10^9$& $5$ & $3.80 \times 10^{1}\,ms$  & $5.19 \times 10^{-1}\,ms$ \\
            $10^9$& $10$ & $5.39 \times 10^{1}\,ms$  & $5.25 \times 10^{-1}\,ms$ \\
            $10^9$& $15$ & $6.72 \times 10^{1}\,ms$  & $6.03 \times 10^{-1}\,ms$ \\
            $10^9$& $20$ & $7.73 \times 10^{1}\,ms$  & $5.82 \times 10^{-1}\,ms$ \\
    \end{tabular}
    \caption{Benchmark times for free size sampling of strict partitions into squares}
    \label{tab:benchstrictsquare}
    \end{table}

\end{example}

\section{Extensions and generalisation} \label{sec:extension}
This section shows how the thinning argument underlying Proposition~\ref{prop:algo} can be reused when $a_n$ is unbounded, provided that one can construct an explicit dominating intensity. The resulting samplers have exactly the same structure as Algorithm~\ref{algo:Boltzmann1}, with only the dominating bound and the proposal distribution for $n$ changing.

\paragraph{Exponential growth bound.}
Assume that the counting sequence satisfies
\begin{equation*}
    a_n \leq bc^n.
\end{equation*}
Then
\begin{equation*}
    \lambda = \sum_{n=0}^\infty a_n\ln(1+z^n) \leq \sum_{n=0}^\infty bc^n z^n=\frac{b}{1-cz}=\overline{\lambda},
\end{equation*}
and we may run Algorithm~\ref{algo:Boltzmann1} with
\begin{equation*}
    \overline{\lambda}=\frac{b}{1-cz},\qquad n\sim \mathrm{Geom}(1-cz),\qquad \text{accept with probability }\ \frac{a_n}{bc^n}\frac{\ln(1+z^n)}{z^n}.
\end{equation*}
This case includes, for instance, classes such as words over an alphabet, random walks, or unlabelled trees.

\paragraph{Linear growth bound and pointing.}
Assume that
\begin{equation*}
    a_n \leq bn.
\end{equation*}
Then
\begin{equation*}
    \lambda = \sum_{n=0}^\infty a_n\ln(1+z^n) \leq \sum_{n=0}^\infty bn z^n = \frac{bz}{(1-z)^2} = \overline{\lambda}.
\end{equation*}
Here the proposal distribution for $n$ is replaced by the distribution with probability mass function
\begin{equation*}
    p_n = nz^{n-1}(1-z)^2, ~~n\in \NN,
\end{equation*}
which we denote by $\mathrm{Geom}_\bullet(1-z)$. The sampler again follows Algorithm~\ref{algo:Boltzmann1}, with
\begin{equation*}
    \overline{\lambda}=\frac{bz}{(1-z)^2},\qquad n\sim \mathrm{Geom}_\bullet(1-z),\qquad \text{accept with probability }\ \frac{a_n}{bn}\frac{\ln(1+z^n)}{z^n}.
\end{equation*}
This case is suitable when a pointing operation, as defined in \cite{anacomb}, is applied to a structure with a bounded counting sequence. If needed, pointing can be iterated in order to consider counting sequences $a_n$ that grow polynomially. In practice it is easy to simulate $Z\sim\mathrm{Geom}_\bullet(1-z)$ as we have
\begin{equation*}
    Z = X+Y+1,
\end{equation*}
where $X$ and $Y$ are mutually independent and follow the distribution $\mathrm{Geom}(1-z)$.

\begin{example}\label{ex:inflection}
    We apply the linear-growth extension of Algorithm~\ref{algo:Boltzmann1} (i.e. $a_n\le bn$) to $\PSet(\NN_\bullet)$. Since the counting sequence of $\NN_\bullet$ is $a_n=n$, we have $b=1$. This can be viewed as integer partitions where pointing distinguishes one unit in each part; we represent this by colouring the parts of the Young diagram as in Figure~\ref{fig:pointing}.

    After rescaling, the empirical profile shows an initial concave segment, an inflection, and a long tail of small parts. By contrast, many classical limit-shape results yield convex profiles with $\sqrt{N}$ scaling on both axes (e.g. \cite{bogachev,comtet,pey}).

    \begin{figure}[h!]
    \centering
    \begin{minipage}{0.49\textwidth}
        \centering
        \includegraphics[width=\linewidth]{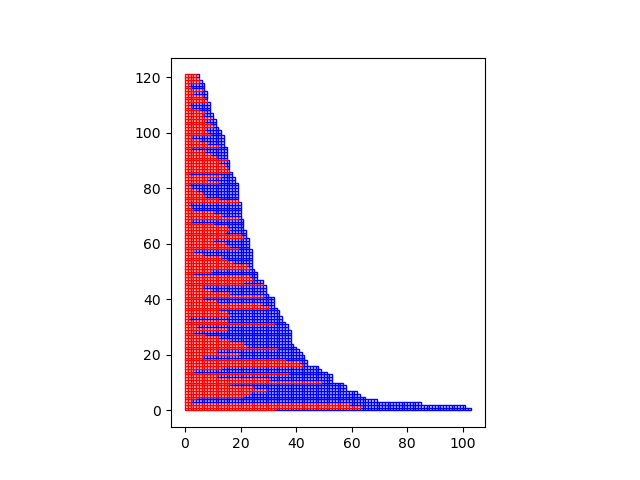}
    \end{minipage}\hfill
    \begin{minipage}{0.49\textwidth}
        \centering
        \includegraphics[width=\linewidth]{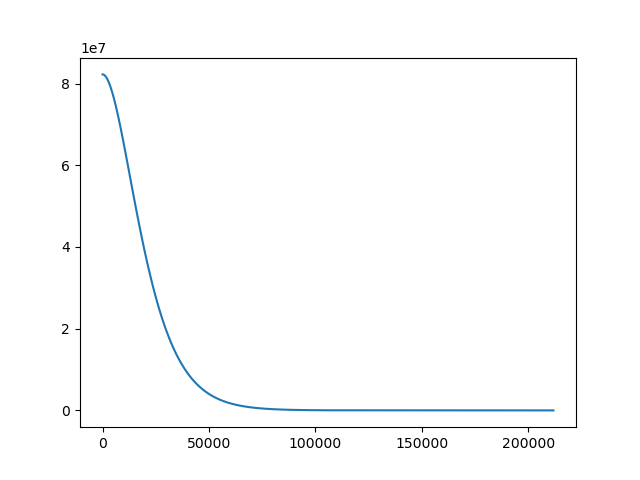}
    \end{minipage}
    \caption{Left: diagram of a sample of $\PSet(\NN_{\bullet})$ of size $N=3,515$. Right: upper boundary for a sample with $N\approx 1.9\times 10^{12}$.}
    \label{fig:pointing}
    \end{figure}
\end{example}

The examples suggest the following general approach. Assume that
\begin{equation*}
    \lambda = \sum_{n=0}^{\infty} a_n \ln(1+z^n) = \sum_{n=0}^{\infty} \lambda_n,
\end{equation*}
and suppose that we can construct a dominating sequence $\lambda_n \leq \overline{\lambda}_n$ such that the sum
\begin{equation*}
    \overline{\lambda} = \sum_{n=0}^{\infty} \overline{\lambda}_n
\end{equation*}
can be evaluated. This leads to the following generic sampling scheme.
\begin{enumerate}
    \item Sample $\overline{M} \sim \mathrm{Poiss}(\overline{\lambda})$.

    \item For each $i\in\{1,\dots,\overline{M}\}$:
    \begin{enumerate}
        \item sample a size $n$ with probability
            $\overline{\lambda}_n/\overline{\lambda}$;

        \item accept the size with probability $\lambda_n/\overline{\lambda}_n$;

        \item if accepted, select an object uniformly in $\Aa_n$.
    \end{enumerate}
\end{enumerate}
The practical implementation relies on constructing a suitable
dominating sequence $(\overline{\lambda}_n)_{n\geq 0}$ such that the sum $\overline{\lambda}$ can be evaluated and the distribution $\overline{\lambda}_n/\overline{\lambda}$ can be sampled efficiently. Finally, uniform sampling in the classes $\Aa_n$ should be feasible and efficient.
    
\section{Discussion}
We have constructed a Boltzmann sampler for $\Cc = \PSet(\Aa)$ that does not require oracle evaluation of the generating function $C(z)$.

While we mainly focused on the case of bounded counting sequence, the same thinning template extends to settings where $a_n\to\infty$, provided an explicit dominating intensity can be constructed. In such cases, uniform sampling in the sets $\Aa_n$ may itself require non-trivial algorithmic work and can be interpreted as a form of ``oracle'' that concerns only the base class $\Aa$ (through $(a_n)$ and uniform sampling in $\Aa_n$), rather than the derived powerset class $\PSet(\Aa)$ or its generating function. Contrary to the sampling scheme of \cite{Flajolet}, Algorithm~\ref{algo:Boltzmann1} is suitable when $\Aa$ contains elements of size zero.

An alternative oracle-free approach was proposed in \cite{pey2}, based on sequential generation with a truncation threshold; deriving such a threshold is case-specific and requires an additional tail analysis.

Finally, Examples~\ref{ex:plain} and \ref{ex:square} indicate that the runtime of Algorithm~\ref{algo:Boltzmann1} is comparable to existing Boltzmann samplers \cite{Flajolet, pey2}, while remaining straightforward to implement.

\section*{Declaration of AI-assisted technologies}
The author used ChatGPT to assist with formatting, language polishing and to correct typographical errors. The author reviewed and edited the manuscript after using this tool and takes full responsibility for the content of the article.

\bibliography{references}

\end{document}